\documentclass[10pt,a4paper]{article}

\usepackage{graphicx}

\usepackage[margin=2.5cm]{geometry}

\begin{document}

\setlength{\parindent}{0em}
\newcommand{\nind}{\setlength{\parindent}{0em}}
\newcommand{\ind}{\setlength{\parindent}{2em}}

{\sf \LARGE Identifying signatures of photothermal current in a double-gated semiconducting nanotube}

\vspace{0.5em}

G. Buchs$^{1,*}$, S. Bagiante$^2$, G. A. Steele$^{3,*}$

\vspace{1em}

{$^1$Centre Suisse d'Electronique et de Microtechnique (CSEM), Jaquet-Droz 1, 2002 Neuch\^{a}tel, Switzerland.} 

{$^2$Laboratory for Micro- and Nanotechnology, Paul Scherrer Institut, 5232 Villigen PSI, Switzerland.} 

{$^3$Kavli Institute of NanoScience, Delft University of Technology, PO Box 5046, 2600 GA, Delft, The Netherlands.

\emph \tiny $^*$Correspondence and requests for materials should be address to
  G.A.S. and/or G.B. 
  
 (email: g.a.steele@tudelft.nl; gilles.buchs@csem.ch).

\vspace{1em}

{\bf 
The remarkable electrical and optical properties of single-walled carbon nanotubes (SWNT) 
have allowed for engineering device prototypes showing great potential for applications such 
as photodectors and solar cells. However, any path towards industrial maturity requires 
a detailed understanding of the fundamental mechanisms governing the process of photocurrent 
generation. Here, we present scanning photocurrent microscopy measurements on a double-gated 
suspended semiconducting SWNT and show that both photovoltaic and photothermal mechanisms 
are relevant for the interpretation of the photocurrent. We find that the dominant or 
non-dominant character of one or the other processes depends on the doping profile, and 
that the magnitude of each contribution is strongly influenced by the series resistance 
from the band alignment with the metal contacts. These results provide new insight into 
the interpretation of features in scanning photocurrent microscopy and lay the foundation 
for the understanding of optoelectronic devices made from SWNTs.}

\pagebreak

\section*{Introduction}

A better understanding of charge transport mechanisms in nanoscale devices, especially the 
role of electrical contact interfaces and band bending, has been made possible in the last 
decade with the development of scanning photocurrent microscopy (SPCM), a dedicated local 
probe technique exploiting local electrical currents generation from light absorption. This 
technique has been used for characterizing various systems like \emph{e.g.} Si nanowires 
\cite{Ahn_2005,Koren_2009,Allen_2009}, colloidal quantum dots \cite{Prins_2012}, VO$_{2}$ 
nanobeams \cite{Kasirga_2012}, carbon nanotubes 
\cite{Balasubramanian_2005,Ahn_2007,Lee_SPCM_2007,Freitag_SPCM_2007,Tsen_2008,Gabor_2009,St-Antoine_2009,St-Antoine_2012,Buchs_JAP} 
as well as 2D materials like graphene \cite{Mueller_2009,Lee_2008,Lemme_2011,Peters_2010,Xia_2009,Park_Ahn_2009,Gabor_2011,Herring_2014} 
and MoS$_{2}$ \cite{Wu_2013,Buscema_2013}. In such nanoscale systems, two mechanisms have 
been identified for the generation of photocurrent: i) photovoltaic processes where 
photo-excited carriers are separated by built-in electrical fields and ii) photothermal 
processes where thermoelectric forces drive carriers through light-induced thermal gradients. 

\setlength{\parindent}{2em}

Single-walled carbon nanotubes (SWNT) are long, one dimensional conductors with a band 
structure ranging from quasimetallic  (bandgap $E_\mathrm{g}\sim $~30 meV) to semiconducting 
character ($E_\mathrm{g}$ typically in the range of 0.1 to 1 eV) depending on chirality and 
diameter. These properties make SWNTs an ideal platform for exploring photocurrent generation 
with SPCM. In early single-walled nanotubes SPCM work the interpretation of photocurrent 
was mostly based on photovoltaic mechanisms 
\cite{Balasubramanian_2005,Ahn_2007,Lee_SPCM_2007,Freitag_SPCM_2007,Gabor_2009}. The 
importance of photothermal effects has been suggested in the context of measurements 
of bulk SWNT films \cite{St-Antoine_2009,St-Antoine_2012,Nanot_2013} and very recently, SPCM work 
on graphene and individual metallic SWNTs has emphasized the importance of photothermal 
mechanisms in materials with no or small bandgaps \cite{Gabor_2011,Barkelid_2013}. The 
question of the role of photothermal mechanisms in larger bandgap semiconducting nanotubes 
has been studied very recently in double-gated \cite{Barkelid_2013} and single-gated 
\cite{DeBorde_2014} suspended carbon nanotube devices. These two studies report contradictory 
results, leaving the understanding of fundamental mechanisms underlying photocurrent 
generation in semiconducting nanotubes unclear.

\setlength{\parindent}{2em}
 
Here we report on the study of a suspended semiconducting nanotube device where we show 
that both photovoltaic and photothermal mechanisms compete in the generation of photocurrent. 
In particular, we find that the dominant or non-dominant character of one or the other 
processes is a function of the doping profile and that the magnitude of each contribution 
is strongly influenced by the band alignment with the metal contacts through the resulting 
contact resistance.

\section*{Results}

{\bf Description of the device. } 
The device consists of a suspended nanotube grown between platinum electrodes over a 
predefined 3 $\mu$m wide and 1 $\mu$m deep trench with four gates defined at the bottom
(see Methods). A schematic and a scanning electron microscopy image of a typical device 
used in this study are shown in Figs. 1(a) and 1(b), respectively.  In our SPCM experiments, 
gate pairs $\mathrm{G1}$-$\mathrm{G2}$ and $\mathrm{G3}$-$\mathrm{G4}$ are connected to independent tunable voltage sources 
$V_\mathrm{G1-G2}$ and $V_\mathrm{G3-G4}$, respectively and source-drain voltage $V_\mathrm{SD}$ 
is set to 0 V. All measurements are performed at atmospheric pressure and room temperature. 
A transistor characteristic curve of the device recorded by sweeping all gate voltages 
simultaneously with $V_\mathrm{SD}=$ 1 mV is displayed in Fig. 1(c). A large conductance 
for holes ($R\approx 300 \mathrm{k}\Omega$) and a very low conductance for electrons ($R > \mathrm{G}\Omega$) 
are observed. From the largely suppressed conductance when the device is pinched-off, we 
conclude that the nanotube is semiconducting with a relatively large bandgap of the order 
of a few to several hundreds of meV \cite{Buchs_JAP}, consistent with the diameter distribution 
expected from the growth recipe used \cite{recipe_Kong}.

\vspace{1em}
\setlength{\parindent}{0em}

{\bf SPCM Measurements. }
Photocurrent (PC) images and corresponding qualitative band diagrams for p-n, n-n, and p-p doping 
configurations are shown in Fig. 2. For n-n and p-n (n-p) configurations, strong PC 
signatures observed at locations corresponding to local depletion regions show signs
which are consistent with photovoltaic processes \cite{Buchs_JAP,Ahn_2007}. The PC image 
for the p-p configuration shows a more complex pattern with alternating sign changes 
along the nanotube axis. The sign of PC spots at the drain and source electrode edges is 
consistent with the photovoltaic mechanism. However, the sign of the PC patterns observed 
inside the trench cannot be explained even qualitatively by photovoltaics effects, since 
it indicates that electrons are travelling ``uphill" along the electrostatic potential 
profile. Similar observations have recently been reported for single gated devices and 
have been attributed to photothermoelectric effects \cite{DeBorde_2014}.

\setlength{\parindent}{2em}

In order to explore the possibility of the presence of such photothermoelectric 
effects in our device, we studied the PC generation mechanisms by inducing 
non-uniform charge density profiles in the suspended nanotube channel using the 
two separate gate pairs. For this, we recorded 2D maps of the PC as a function of 
the gate pairs voltages ($V_\mathrm{G1-G2}$ and $V_\mathrm{G2-G3}$) for a fixed 
location of the laser spot along the nanotube axis. The resulting 2D map for the 
laser spot located at the center of the suspended portion of the SWNT is shown 
in Fig. 3(a). Four doping regions, clockwise p-n, n-n, n-p and p-p, are identified 
across the gates voltages space and are delimited by two dashed black lines. A first 
clear feature of the image is a strongly suppressed PC in the n-n region. This can 
be easily understood from the very high contact resistance ($R > \mathrm{G}\Omega$) measured 
in this regime (Fig. 1(c)). The p-n and n-p regions show PC signatures with a sign 
that is consistent with the photovoltaic mechanism (Note that the slight clockwise 
tilt angle of the strip shaped pattern is most probably due to the fact that the laser 
spot is not exactly centred on the n-p (or p-n) depletion zone but slightly shifted 
towards the drain contact. This has no incidence on the main conclusions of our present work). 
However, the p-p region shows PC features which are not consistent with a purely photovoltaic 
mechanism. This can be verified if one considers the line trace in Fig. 3(b) 
corresponding to a transition from p-n to p-p$^{+}$ doping configuration along 
the vertical dashed line in the PC map, which reveals two sign reversals of the 
PC signal. From a purely photovoltaic mechanism, one would expect a monotonic 
increase of the PC signal from the p-n minimum to a positive value in the p-p$^{+}$ 
region with a zero crossing at the symmetric p-p doping profile. This would show 
only one single PC sign reversal corresponding to a sign reversal of the electric 
field at the laser spot location. The exact same analysis applies for the PC signal 
recorded along the horizontal dashed line in the 2D map (Fig. 3(c)).

\vspace{1em}
\setlength{\parindent}{0em}

{\bf Model for photothermal currents. }
From the above analysis, it is clear that the photocurrent in the p-p region of 
Fig. 3 cannot be described by a photovoltaic mechanism, even at a qualitative level. 
In Fig. 4, we show how the observations in Fig. 3 can be explained with a simple model 
where the suspended nanotube is separated in two regions of different carriers density 
on the left and right side of the laser spot. The model also includes the effect of 
laser heating at the junction between the two regions. The left (drain contact) and 
right (source contact) portions have average Seebeck coefficients $S_\mathrm{1}$ and 
$S_\mathrm{2}$, respectively. The laser light induces a temperature increase $\Delta T$ 
with respect to the contacts which can be regarded as heat sinks. We consider linear 
temperature profiles from the $S_{1}-S_{2}$ interface to the contacts 
\cite{DeBorde_2014,St-Antoine_2009}. Photothermal currents are induced by the local 
electromotive field $\mathbf{E}_\mathrm{PT}=-S \nabla T$ where $\nabla T$ is the 
temperature gradient along the nanotube axis and $S$ is the Seebeck coefficient. 
Thus the generated photothermal current can be expressed by $I_\mathrm{PT}=-R^{-1}\cdot \int S(x) \nabla T dx$ 
where $R$ is the overall contact resistance. Applied to our thermocouple model, we find:
	\begin{equation}
	I_\mathrm{PT}=R^{-1}\cdot \vert \Delta T \vert \cdot \left(S_\mathrm{2}-S_\mathrm{1} \right)
	\end{equation} 
The sign of the above expression (conventional flow notation) is then given by the Seebeck 
coefficients , whose absolute values and signs depend on the doping levels. A qualitative 
profile of the Seebeck coefficient as a function of the Fermi level for a semiconducting 
SWNT is shown in Fig. 4(a) \cite{Small_2003,DeBorde_2014}. 

\vspace{1em}
\setlength{\parindent}{0em}

{\bf Analysis and interpretation of the experimental results. }
Band diagrams corresponding to doping configurations along the vertical line in Fig. 3(a), 
\emph{i.e.} p-n, p-p$^{-}$ and p-p$^{+}$ are drawn in Figs. 4(c)-(e), respectively. The 
photothermal and photovoltaic currents are indicated with arrows oriented with respect to 
the electronic flow notation (colors correspond to the current sign with respect to the 
conventional flow notation, \emph{i.e.} red for positive and blue for negative). From this, 
we find that the photovoltaic and photothermal currents in the case of the p-n (and n-p) 
configuration flow in the same direction. However, we find that photovoltaic and photothermal 
currents are flowing in opposite directions for p-p$^{-}$ and p-p$^{+}$ (p$^{-}$-p and p$^{+}$-p). 
To understand these opposite flowing currents, it is important to contrast the mechanisms 
driving the two. For photovoltaic currents, electron flow is driven by differences in 
electrostatic voltage: electrons move to the point of lowest energy. For photothermal 
currents, electrons are driven by differences in chemical potential and in some cases can 
cause electrons to flow ``uphill``. The photothermal current signs given by our model for 
gates voltages configurations corresponding to a p-doped nanotube channel (Fig. 4(d)-(e)) 
are consistent with the experimental results (p-p region in Fig. 3(a)), indicating that 
photothermoelectric effects are dominant in our device in this regime. From equation 1 and 
the maximum PC in the p-p region of about 20 pA corresponding to a Seebeck coefficient 
difference $S_{2}-S_{1} \approx$ 140 $\mu$V/K \cite{Small_2003}, we estimate the laser-induced 
temperature increase $\Delta T$ to be in the order of $\sim$ 50 mK. This is smaller than what 
was reported in a previous work [26]. This could be a result of the approximation of an abrupt 
junction in our model, from different net laser power reaching the sample, or from a different 
ratio between excitation energy and optical resonances of the nanotube.

\setlength{\parindent}{2em}

From the data in Fig. 3, it is also quite clear that the largest photocurrent occurs in the 
p-p region, suggesting that photothermal mechanisms are the most dominant in our device for 
all gate voltages. However, we note that one has to be careful in comparing photocurrent levels 
from different doping configurations of the device since the photocurrent is also sensitive to 
series resistances to the contacts. A more relevant comparison figure for different doping 
levels is the photo-induced voltage: for our device, in the p-n configuration, the photovoltage 
should be quite large, on the order of the band gap ($\sim$ 100 meV). In contrast, in the p-p 
regions where the photocurrent is the largest, the photothermal voltage estimated from the device 
resistance ($R\sim$ 300 k$\Omega$) and observed current is on the order of $\approx$ 10 $\mu$ eV. 
Although the photovoltage in the p-n and n-p regions should be significantly larger than in the 
p-p region, the photocurrent is comparable or smaller due to the very large n-type contact series 
resistance ($R > \mathrm{G}\Omega$).

\setlength{\parindent}{2em}

In previous studies with graphene \cite{Gabor_2011,Levitov_2011}  and metallic nanotubes 
\cite{Barkelid_2013}, 2D gate maps such as the one in Fig. 3(a) showed a very characteristic 
6-fold rotational symmetry pattern that was used as a fingerprint for identifying a dominating 
photothermal effect. In contrast, our semiconducting suspended nanotube shows instead a more 
4-fold-like pattern. This lack of symmetry finds its origin in the fact that we are working 
with a semiconducting material with asymmetric n- and p-type contact resistances. In particular, 
the suppressed PC signal in the n-n region is due to the very large n-type contact resistance 
($R > G\Omega$). For graphene (a semi-metal) and metallic SWNTs, the lack of an energy bandgap 
allows for similarly low p-type and n-type contact resistances, resulting in equal intensity PC 
signatures both in p-p and n-n regions of the 2D maps, thus forming a 6-fold pattern.

\setlength{\parindent}{2em}

Furthermore, we note that our conclusion for dominant photothermal PC in the p-type regime
is in contradiction with a recent work \cite{Barkelid_2013} in which it was suggested that 
the photovoltaic mechanism dominates photocurrent generation in suspended semiconducting 
nanotubes for all doping configurations. We address these apparent contradictory results by 
noting that compared to our device, the devices in Ref. \cite{Barkelid_2013} had a much lower 
contact resistance for n-type doping ($\sim$ 100 M$\Omega$) and a much higher resistance for 
p-type doping ($\sim$10 M$\Omega$). This suggests a more symmetric effective work function 
alignment in their devices, with the Fermi level of the leads pinned near the mid-gap level 
(this difference in work function alignment is likely due to the vacuum environment and the 
vacuum thermal annealing procedure reported, which likely removed an adsorbed water layer 
on the surface of the silicon oxide in the trench). With the larger Schottky barriers for 
p-type doping in Ref. \cite{Barkelid_2013} the photothermal currents are likely suppressed 
to a level close to or below the noise floor of the measurement by the large series resistance 
to the contacts. 

\section*{Discussion}

In summary, we have shown that for the interest of interpreting the origin of photocurrent 
features in SPCM with semiconducting nanotubes, photothermal effects can absolutely not be 
a-priori excluded. Similar to systems like graphene and metallic SWNTs, suspended semiconducting 
SWNTs can show strong photothermoelectric effects through measurement of photothermal currents 
induced by inhomogeneous doping profiles along thermal gradients. However, the dominant or 
non-dominant character of photothermal currents compared with photovoltaic currents strongly 
depends on the doping profile of the device and on the n- and p-type contact resistances. 

\section*{Methods}

{\bf Sample fabrication. }
The fabrication of our device began with a p++ silicon wafer used as a backgate covered by 
285 nm of thermal silicon oxide. On top of this, gate electrodes made of 5/25 nm W/Pt were 
defined using electron-beam lithography, followed by the deposition of a 1100 nm thick SiO$_{2}$ 
layer. A 1000 nm deep trench was dry etched, leaving a thin oxide layer on top of the gates. 
A 5/25 nm W/Pt layer was then deposited to serve as source and drain contacts, and a single-walled 
carbon nanotube was grown at the last fabrication step at a temperature of 900 $^{\circ}{\rm C}$ 
from patterned Mo/Fe catalysts \cite{Steele_2009,recipe_Kong}.

\vspace{1em}
\setlength{\parindent}{0em}

{\bf SPCM setup. }
Our SPCM setup \cite{Buchs_JAP} consists of a confocal microscope system with the 
objective ($NA=$ 0.8) illuminated by $\lambda=$ 532 nm continuous wave collimated 
laser light. The diffraction limited spot with a measured diameter of $\sim$ 330 nm 
is scanned across the sample surface by means of two orthogonal galvo-mirrors (x,y) 
combined to a telecentric lens system while the PC signal as well as the reflected 
light intensity are recorded simultaneously in order to determine the 
absolute position of the detected PC features. 

\vspace{1em}
\setlength{\parindent}{0em}

{\bf Measurements. }
All measurements presented in this work are performed at atmospheric pressure and 
room temperature. For SPCM measurement, the source-drain voltage is maintained to 0 V 
and typical light intensities of 5 kW/cm$^{2}$ are used.

\section*{Acknowledgments}

The authors acknowledge Val Zwiller for experimental support. 
This research was supported by a Marie Curie Intra European Fellowship within the 7th European 
Community Framework Programme (G.B.) and the Netherlands Organization for Scientific Research, 
NWO (G.A.S.) and the University of Catania (S.B.).

\section*{Author contributions}

G.B. and S.B. performed the experiments; S.B. fabricated the sample; G.B. and G.A.S wrote the 
manuscript; all authors discussed the results and contributed to the manuscript.

\vspace{1em}

The authors declare that they have no competing financial interests.

\pagebreak

\begin{figure}
\begin{center}
\includegraphics[]{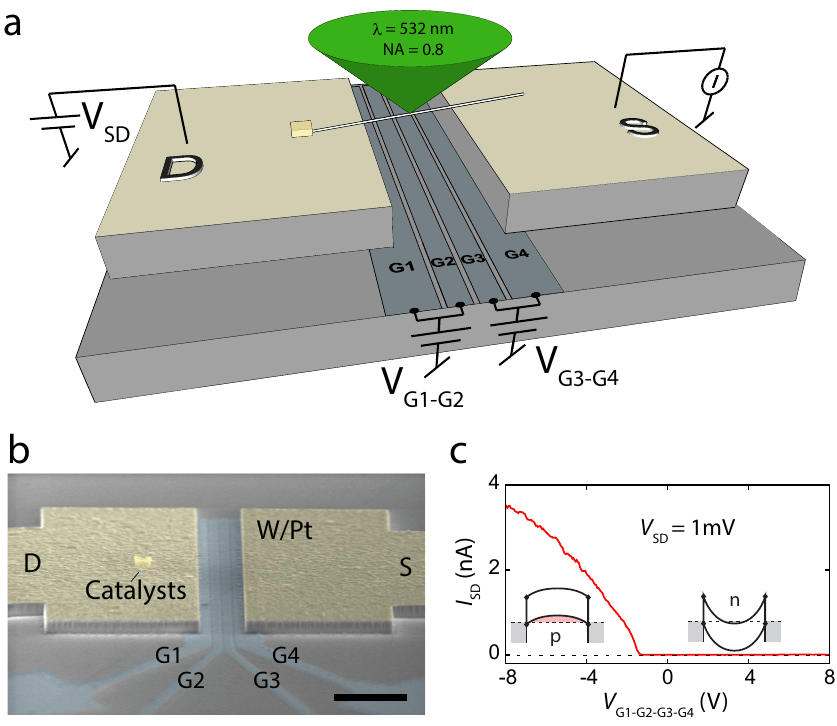}
\caption{\textbf{Description and characterization of the device. } \textbf{a}) Schematic of 
the device. Gates G1-G2, G3-G4 are connected to tunable voltage sources $V_\mathrm{G1-G2}$ 
and $V_\mathrm{G2-G3}$, respectively. The diffraction-limited laser spot ($\lambda = $532 nm) 
is accurately positioned along the suspended nanotube axis and induced photocurrent is recorded 
at the source (S) contact. (\textbf{b}) Scanning electron microscope image of a typical device. Scale bar: 5 $\mu$m. (\textbf{c}) 
Transistor curve of the device recorded by sweeping gate voltages $V_\mathrm{G1-G2-G3-G4}$ 
simultaneously with $V_\mathrm{SD}=$ 1 mV. Onsets: qualitative band diagrams in p- and n-doping 
regimes, corresponding to "ON" and "OFF" states, respectively. }
\end{center}
\end{figure}

\pagebreak

\begin{figure}
\begin{center}
\includegraphics[]{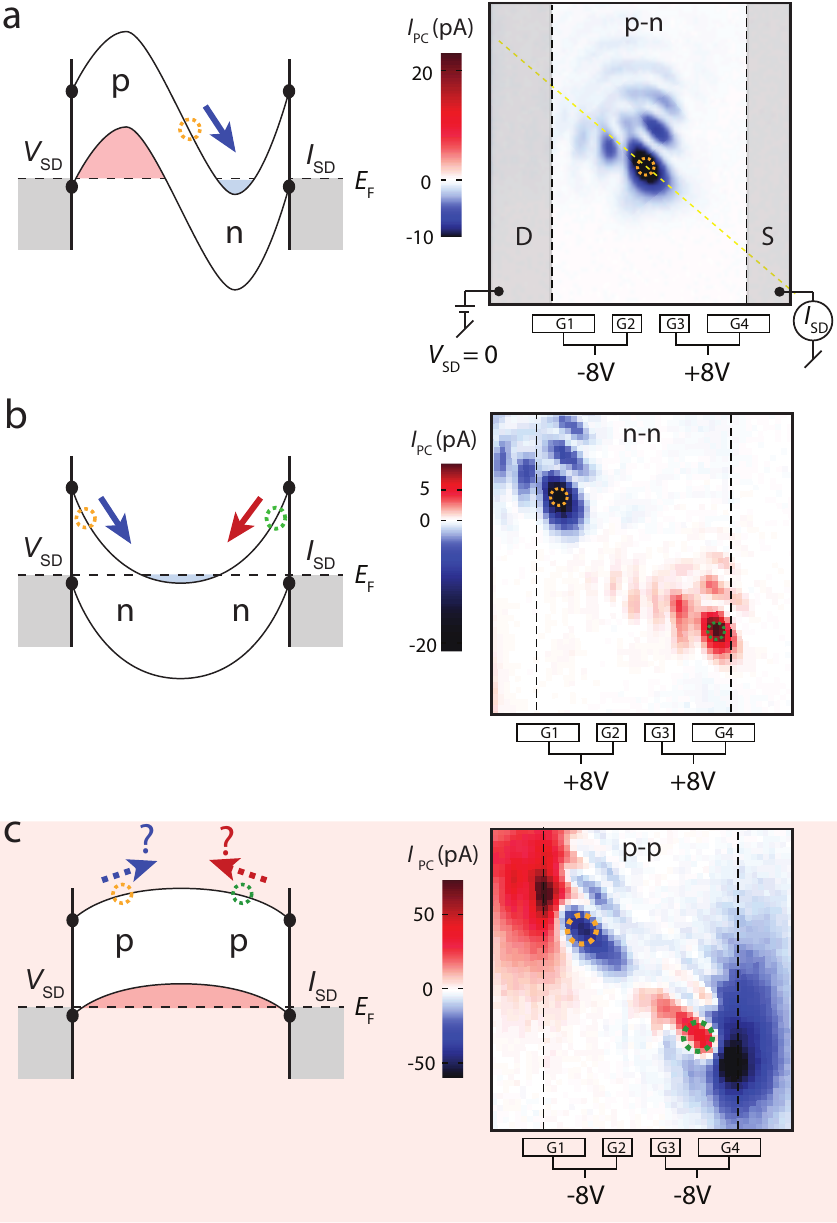}
\caption{\textbf{Photocurrent images for different doping configurations. }(\textbf{a})-(\textbf{c}) 
Qualitative band diagrams (left) and photocurrent images (right) measured for p-n, n-n, and p-p doping 
configurations, respectively. Location of trench edges (vertical black dashed lines) and gates with 
applied voltages are indicated. Source ($S$) and drain ($D$) contacts with corresponding electrical 
connections are highlighted in (a) and a yellow dashed line shows the position of the nanotube axis. 
The location of photocurrent extrema are indicated by dashed circles. The sign of the photocurrent 
corresponds to the conventional flow notation (red: positive, blue: negative). The arrows on the band 
diagrams are oriented according to the electron flow notation. Panel (c) illustrates the dominant 
photothermal mechanism observed in the p-p configuration where electrons are travelling ``uphill`` 
(dashed arrows in the band diagram). Patterns around the maximum intensity PC spots in the PC images 
are due to diffraction effects at the trench gates \cite{Buchs_JAP}. }
\end{center}
\end{figure}

\pagebreak

\begin{figure}
\begin{center}
\includegraphics[]{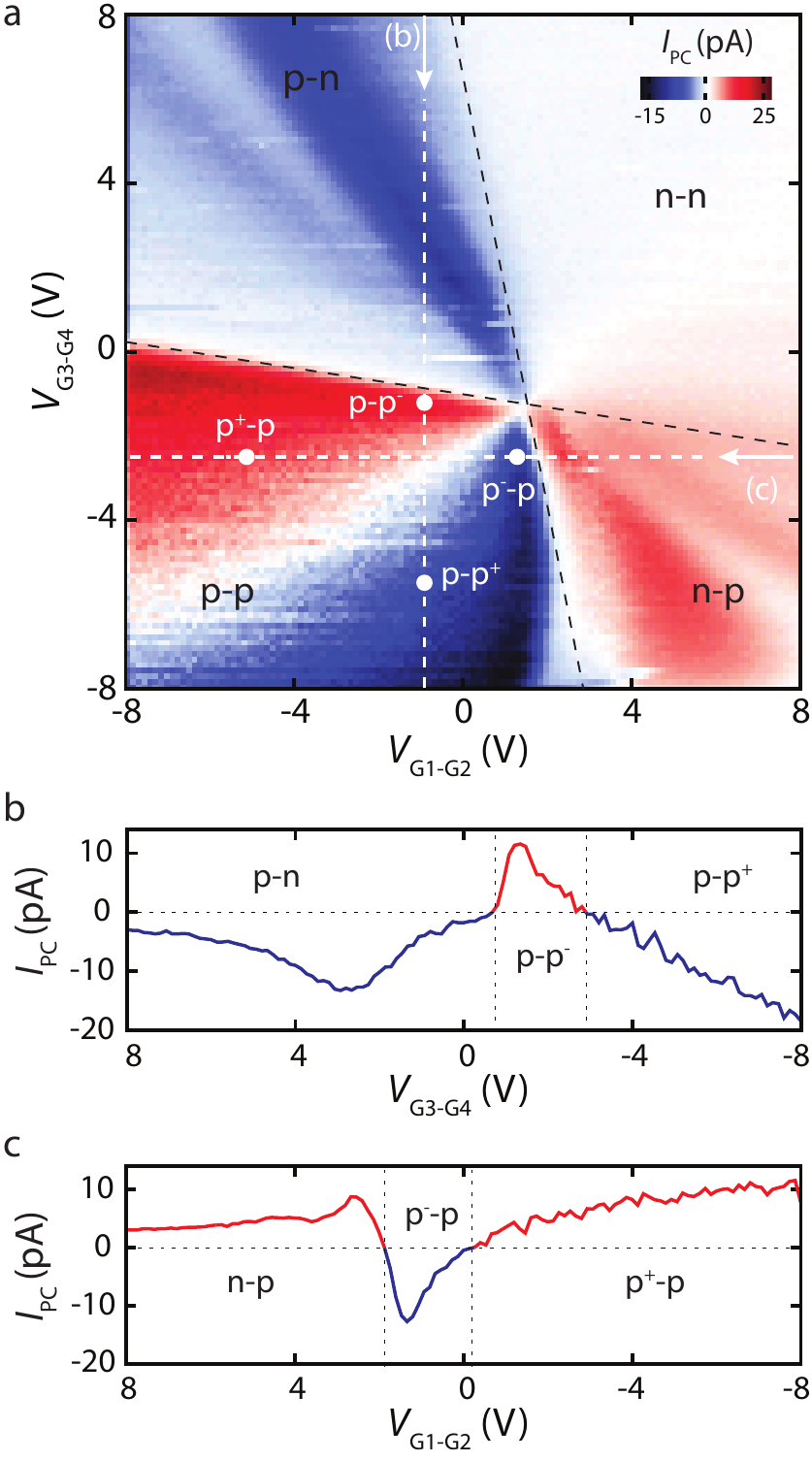}
\caption{\textbf{Gate dependence of the photocurrent demonstrating a clear dominating photothermal 
response. }(\textbf{a}) 2D map of the measured photocurrent ($I_\mathrm{PC}$) versus $V_\mathrm{G1-G2}$ 
and $V_\mathrm{G3-G4}$ recorded with laser focussed near the middle section of the suspended nanotube, 
close to the p-n photocurrent minimum shown in Figure 2(a). Black dashed lines delimit four p-n, n-n, 
n-p and p-p doping regions. (\textbf{b}) and (\textbf{c}) $I_\mathrm{PC}$ line traces recorded along 
the horizontal, respectively vertical white dashed lines in panel (a). n-p, p$^{-}$-p and p$^{+}$-p 
(p-n, p-p$^{-}$ and p-p$^{+}$) doping regions are highlighted.
}
\end{center}
\end{figure}

\pagebreak

\begin{figure}
\begin{center}
\includegraphics[]{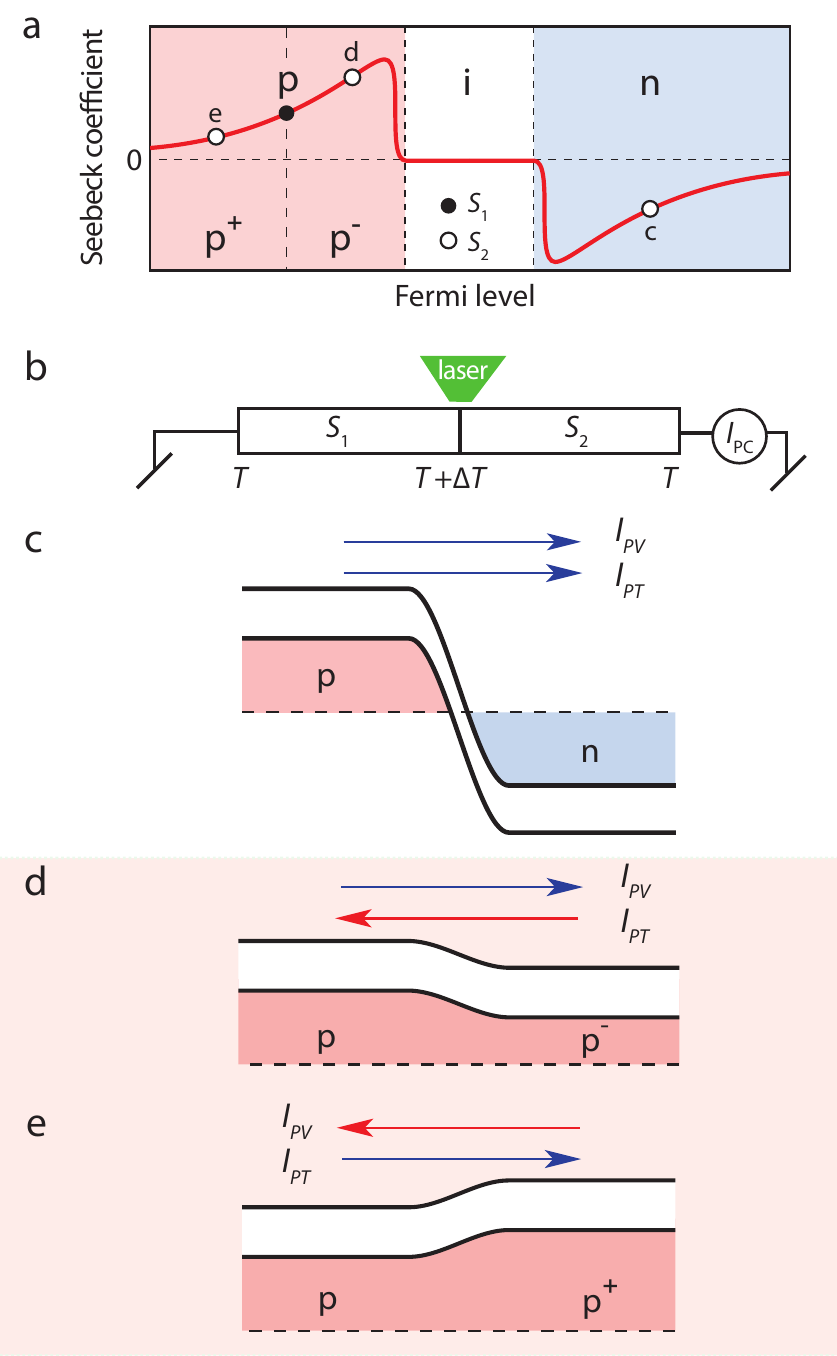}
\caption{\textbf{A simple model for photothermal currents travelling against an electric field. }
(\textbf{a}) Qualitative behavior of the Seebeck coefficient of a semiconducting carbon nanotube 
versus Fermi level. (\textbf{b}) Thermocouple model with two separately gated nanotube regions 
with corresponding Seebeck coefficients $S_\mathrm{1}$ and $S_\mathrm{2}$. Laser excitation region 
located at the two regions interface gives rise to a temperature increase $\Delta T$. (\textbf{c})
-(\textbf{e}) Band diagrams for p-n, p-p$^{-}$ and p-p$^{+}$ doping configurations with corresponding 
photovoltaic and photothermal current arrows oriented with respect to the electronic flow notation. 
The color corresponds to the current sign with respect to the conventional flow notation.
}
\end{center}
\end{figure}

\end{document}